\documentclass[10pt,aps,prd,showpacs,showkeys,amssymb,preprintnumbers,nofootinbib,
superscriptaddress,notitlepage,eqsecnum]{revtex4}

\usepackage{amsmath}
\usepackage{latexsym}
\usepackage[usenames]{color}
\usepackage{amsfonts}
\usepackage{url,hyperref}
\usepackage{bm}

\hypersetup{colorlinks=true, citecolor=blue}
\definecolor{dred}{rgb}{0.75,0.08,0}

\usepackage{graphics}

\newcommand{\nn}{\nonumber\\}
\newcommand{\beq}{\begin{equation}}
\newcommand{\eeq}{\end{equation}}
\newcommand{\bed}{\begin{displaymath}}
\newcommand{\eed}{\end{displaymath}}
\def\bea{\begin{eqnarray}}
\def\eea{\end{eqnarray}}

\begin{document}

\title{Scalar spheroidal harmonics in five dimensional Kerr-(A)dS}
\author{H.~T.~Cho}
\email[Email: ]{htcho@mail.tku.edu.tw}
\affiliation{Department of Physics, Tamkang University, Tamsui, Taipei, Taiwan, Republic of 
China}
\author{A.~S.~Cornell}
\email[Email: ]{alan.cornell@wits.ac.za}
\affiliation{National Institute for Theoretical Physics; School of Physics, University of the 
Witwatersrand, Wits 2050, South Africa}
\author{Jason~Doukas}
\email[Email: ]{jasonad@yukawa.kyoto-u.ac.jp}
\affiliation{Yukawa Institute for Theoretical Physics, Kyoto University, Kyoto, 606-8502, Japan}
\author{Wade~Naylor}
\email[Email: ]{naylor@phys.sci.osaka-u.ac.jp}
\affiliation{International College \& Department of Physics, Osaka University, Toyonaka, Osaka 560-0043, Japan}

\begin{abstract}
We derive expressions for the general five-dimensional metric for Kerr-(A)dS black holes. The Klein-Gordon equation is explicitly separated and we show that the angular part of the wave equation leads to just one spheroidal wave equation, which is also that for charged five-dimensional Kerr-(A)dS black holes. We present results for the perturbative expansion of the angular eigenvalue in powers of the rotation parameters up to 6th order and compare numerically with the continued fraction method.
\end{abstract}

\date{\today}
\preprint{YITP-11-63, WITS-CTP-74, OU-HET-708/2011}
\maketitle
\section{Introduction}
\label{intro}

\par Rotating black holes in higher dimensions were first discussed in the seminal paper of Myers and Perry \cite{Myers:1986un}.  One surprising feature of the Myers-Perry (MP) solution is that in general $\lfloor\frac{D-1}{2}\rfloor$ spin parameters are required. The first asymptotically non-flat five-dimensional MP metric was given in \cite{Hawking:1998kw}, where subsequent generalizations to arbitrary dimension, with multiple rotation parameters was done by Gibbons, L\"u, Pope and Page (GLPP) \cite{Gibbons:2004js,Gibbons:2004uw}, which gave a formal proof of the solution in \cite{Hawking:1998kw}. The most general Kerr-(A)dS-NUT metric was found by Chen, L\"u and Pope (CLP) \cite{Chen:2006xh}. These two papers represent a fundamental step to developing separable equations of motion.

\par Deriving wave equations for higher dimensional rotating black hole spacetimes \cite{Frolov:2002xf} relies crucially on the method of separation of variables from a Hamilton-Jacobi equation, see also \cite{Vasudevan:2005js}. Using the CLP metric, Frolov, Krtous and Kubiznak \cite{Frolov:2006pe} were able to separate the geodesic equation and find the Klein-Gordon equation in the most general setting (this was generalized to charged cases in reference \cite{Aliev:2008yk} for zero NUT parameter). The actual separation is due to the presence of hidden symmetries in the form of Killing tensors \cite{Krtous:2006qy, Frolov:2007nt, Frolov:2008jr}, where a whole tower of Killing tensors and symmetry operators \cite{Sergyeyev:2007gf} can be constructed with the help Killing-Yano and conformal Killing-Yano tensors. They guarantee the separability of the geodesic equation, the Klein-Gordon equation and also the Dirac equation \cite{Oota:2007vx}. Unfortunately the separation of the graviton does not appear to be possible based on these hidden symmetries \cite{Frolov:2007nt}.

\par In the literature research has been largely directed toward solutions with only one rotation parameter, the so called simply-rotating case, e.g., see \cite{Kodama:2009rq}. However, recently, solutions with two rotation parameters in $D\geq 6$ have been investigated \cite{Doukas:2010be,Cho:2011yp}. In five dimensions, some work has been performed in the asymptotically flat case \cite{Frolov:2002xf, Nomura:2005mw}, and in this paper we extend these results to non-asymptotically flat cases. These are relevant to the AdS/CFT correspondence, where black hole solutions in five-dimensional minimal gauged supergravity models require rotating solutions to avoid closed timelike curves (CTCs) \cite{Chong:2005hr,Aliev:2008yk,Wu:2009}.  The most general SUGRA solution in five-dimensions for arbitrary rotation parameters ($a_1,a_2$) was found recently in the work of Chong and collaborators \cite{Chong:2005hr}. The important point about the spheroidal harmonics in these cases is that they do not depend on the charge and hence will be the same as that for Kerr-(A)dS.\footnote{Based on symmetry arguments in the metric we expect that the spheroidal harmonics are also independent of NUT charge. Hence our scalar spheroidal harmonics should actually be valid for {\it charged} Kerr-(A)dS-NUT black holes.}

\par This article is a follow up to \cite{Cho:2011yp} in which the Klein Gordon equation for doubly rotating black holes in $D\geq 6$ dimensions was investigated. Superficially,  $D=5$ would not appear to be significantly different from $D\geq6$; however,  it turns out that in five dimensions only a single angular equation occurs, thereby making the analysis different in  these two cases. 

\par In this work we will, after separating the Klein-Gordon equation on the GLPP background (with two rotations in five dimensions) investigate the (scalar) spheroidal harmonics, where in particular we show in detail how to apply perturbation theory to obtain the angular eigenvalues (separation constant) of the spheroidal harmonics. This is in the spirit of Fackerell \& Crossman, and Seidel \cite{Fackerell:1977, Seidel:1988ue}, which used the properties of Jacobi functions. 

\par There are various reasons for studying this case. One is that in five dimensions the rotation parameter is bounded to something like $a_1\omega \leq 1.5$ (for $a_2=0$) in units of black hole mass $M=1$ and hence a perturbative expansion of the eigenvalue is well suited to speed up numerics for QNMs or greybody factors. Furthermore, as we mentioned, because there is only one spheroidal equation for a $D=5$ doubly rotating black hole, the perturbative expansion is much simpler. 

\par The structure of the paper is as follows: In the next section (Section \ref{CLP}), we discuss the general metric for Kerr-(A)dS black holes with two rotations in five-dimensions. The corresponding Klein-Gordon equation is separated into a radial and angular equation in Section \ref{sec:KG}. In section \ref{Ang5} we explain the perturbative method. Then in Section \ref{res} (as a sanity check) we compare our analytics with the continued fraction method.  Conclusions are then given in Section \ref{conc}. 

\section{\bf CLP metric in five dimensions}
\label{CLP}

\par Many of the steps for the separation follow closely to that for $D\geq 6$ doubly rotating black holes \cite{Cho:2011yp}, and hence, we shall only briefly outline the steps. As we are interested in the five dimensional case, we take the CLP metric to be \cite{Chen:2006xh}:
\begin{eqnarray}
ds^{2}&=&\frac{U_{1}}{X_{1}}dy^{2}-\frac{U_{2}}{X_{2}}dr^{2}+\frac{X_{1}}{U_{1}}\left[\frac{\tilde{W}}{1-g^{2}y^{2}}d\tilde{t}-\frac{a_1^{2}\tilde{\gamma}_{1}}{a_1^{2}-y^{2}}d\tilde{\phi}_{1}-\frac{a_2^{2}\tilde{\gamma}_{2}}{a_2^{2}-y^{2}}d\tilde{\phi}_{2}\right]^{2}\nonumber\\
&&\ \ \ +\frac{X_{2}}{U_{2}}\left[\frac{\tilde{W}}{1+g^{2}y^{2}}d\tilde{t}-\frac{a_1^{2}\tilde{\gamma}_{1}}{a_1^{2}+r^{2}}d\tilde{\phi}_{1}-\frac{a_2^{2}\tilde{\gamma}_{2}}{a_2^{2}+r^{2}}d\tilde{\phi}_{2}\right]^{2}\nonumber\\
&&\ \ \ +\frac{a_1^{2}a_2^{2}}{y^{2}r^{2}}\left[\tilde{W}d\tilde{t}-\tilde{\gamma}_{1}d\tilde{\phi}_{1}-\tilde{\gamma}_{2}d\tilde{\phi}_{2}\right]^{2}\;,\label{eqn:1}
\end{eqnarray}
where
\begin{eqnarray}
U_{1}&=&-(y^{2}+r^{2})\;,\\ 
U_{2}&=&y^{2}+r^{2}\;,\\
X_{1}&=&\frac{1}{y^{2}}(1-g^{2}y^{2})(a_1^{2}-y^{2})(a_2^{2}-y^{2})\;,\\
X_{2}&=&-\frac{1}{r^{2}}(1+g^{2}r^{2})(a_1^{2}+r^{2})(a_2^{2}+r^{2})+2M\;,\\
\tilde{W}&=&(1-g^{2}y^{2})(1+g^{2}r^{2})\;,\\
\tilde{\gamma}_{1}&=&(a_1^{2}-y^{2})(a_1^{2}+r^{2})\;,\\
\eea
and
\begin{eqnarray}
t&=&\tilde{t}(1-g^{2}a_1^{2})(1-g^{2}a_2^{2})\;,\\
\phi_{1}&=&\tilde{\phi}_{1}a_1(1-g^{2}a_1^{2})(a_1^{2}-a_2^{2})\;,\\
\phi_{2}&=&\tilde{\phi}_{2}a_2(1-g^{2}a_2^{2})(a_2^{2}-a_1^{2})\;.
\end{eqnarray}
Here we shall only consider the Kerr-(A)dS case, so all NUT charges have been set to zero, where $g^{2}$ is related to the five dimensional cosmological constant with $R_{\mu\nu}=-3g^{2}g_{\mu\nu}$. Moreover, this metric is already in the Boyer-Lindquist form because there is no cross term for $dr$. The ingenuity of the CLP metric is the introduction of the coordinate $y$, which is related to the original direction cosines coordinates. In five dimensions these are $\mu_{1}=\sin\theta$ and $\mu_{2}=\cos\theta$, with
\begin{equation}
\mu_{1}^{2}=\frac{a_1^{2}-y^{2}}{a_1^{2}-a_2^{2}}\ \ \ ,\ \ \
\mu_{2}^{2}=\frac{a_2^{2}-y^{2}}{a_2^{2}-a_1^{2}} \;,
\end{equation}
or
\begin{equation}
y^{2}=a_1^{2}\cos^{2}\theta+a_2^{2}\sin^{2}\theta\;.
\label{lati}
\end{equation}

This metric in equation (\ref{eqn:1}) can be further simplified via the relations \cite{Chen:2006xh}:
\begin{eqnarray}
\frac{\tilde{W}}{1-g^{2}y^{2}}d\tilde{t}-\frac{a_1^{2}\tilde{\gamma}_{1}}{a_1^{2}-y^{2}}d\tilde{\phi}_{1}
-\frac{a_2^{2}\tilde{\gamma}_{2}}{a_2^{2}-y^{2}}d\tilde{\phi}_{2} &=&(d\tilde{t}-a_1^{4}d\tilde{\phi}_{1}-a_2^{4}d\tilde{\phi}_{2})-r^{2} (-g^{2}d\tilde{t}+a_1^{2}d\tilde{\phi}_{1}+a_2^{2}d\tilde{\phi}_{2})\nonumber\\
&\equiv&d\psi_{0}-r^{2}d\psi_{1}\;;
\end{eqnarray}
\begin{equation}
\frac{\tilde{W}}{1+g^{2}y^{2}}d\tilde{t}- \frac{a_1^{2}\tilde{\gamma}_{1}}{a_1^{2}+r^{2}}d\tilde{\phi}_{1} -\frac{a_2^{2}\tilde{\gamma}_{2}}{a_2^{2}+r^{2}}d\tilde{\phi}_{2} = d\psi_{0}+y^{2}d\psi_{1}\;,
\end{equation}
and 
\begin{equation}
\tilde{W}d\tilde{t}-\tilde{\gamma}_{1}d\tilde{\phi}_{1}-\tilde{\gamma}_{2}d\tilde{\phi}_{2} =d\psi_{0}+(y^2-r^2)d\psi_{1}-y^{2}r^{2}d\psi_{2}\;.
\end{equation}
In the above the linear relations are
\begin{eqnarray}
\psi_{0}&=&\tilde{t}-a_1^{4}\tilde{\phi}_{1}-a_2^{4}\tilde{\phi}_{2}\\
\psi_{1}&=&-g^{2}\tilde{t}+a_1^{2}\tilde{\phi}_{1}+a_2^{2}\tilde{\phi}_{2}\\
\psi_{2}&=&g^{4}\tilde{t}-\tilde{\phi}_{1}-\tilde{\phi}_{2}\;,
\end{eqnarray}
or
\begin{eqnarray}
t&=&\psi_{0}+(a_1^{2}+a_2^{2})\psi_{1}+a_1^{2}a_2^{2}\psi_{2}\label{eqn:20}\\
\frac{\phi_{1}}{a_1}&=&\psi_{1}+a_2^{2}\psi_{2}+g^{2}(\psi_{0}+a_2^{2}\psi_{1})\label{eqn:21}\\
\frac{\phi_{2}}{a_2}&=&\psi_{1}+a_1^{2}\psi_{2}+g^{2}(\psi_{0}+a_1^{2}\psi_{1})\;.\label{eqn:22}
\end{eqnarray}
The metric in equation (\ref{eqn:1}) becomes
\begin{eqnarray}
ds^{2}&=&\frac{U_{1}}{X_{1}}dy^{2}-\frac{U_{2}}{X_{2}}dr^{2} +\frac{X_{1}}{U_{1}}\left[d\psi_{0}-r^{2}d\psi_{1}\right]^{2} +\frac{X_{2}}{U_{2}}\left[d\psi_{0}+y^{2}d\psi_{1}\right]^{2}\nonumber\\
&&\ \ \ +\frac{a_1^{2}a_2^{2}}{y^{2}r^{2}}\left[d\psi_{0}+(y^2-r^2)d\psi_{1}-y^{2}r^{2}d\psi_{2}\right]^{2}\;.
\label{diag}
\end{eqnarray}
Note that, comparing with the CLP metric, see equation (22) in reference \cite{Chen:2006xh}, we have
\begin{eqnarray}
&&A_{1}^{(0)}=1\ \ \ ,\ \ \ A_{1}^{(1)}=-r^{2}\ \ \ ,\ \ \
A_{2}^{(0)}=1\ \ \ ,\ \ \ A_{2}^{(1)}=y^{2}\ \ \ ,\nonumber\\
&&A^{(0)}=1\ \ \ ,\ \ \ A^{(1)}=-r^{2}+y^{2}\ \ \ ,\ \ \
A^{(2)}=-y^{2}r^{2}\;.
\end{eqnarray}

\section{\bf Separation of the Klein-Gordon equation}
\label{sec:KG} 

\par Using the results of Frolov, Krtous, and Kubiznak \cite{Frolov:2006pe} and considering their equation (4.2), the Klein-Gordon field in five dimensions has the following ansatz:
\begin{equation}
\Phi=R_{1}(y)R_{2}(r)e^{i\Psi_{0}\psi_{0}+i\Psi_{1}\psi_{1}+i\Psi_{2}\psi_{2}}\;.
\end{equation}
The separated equations for $R_{1}$ and $R_{2}$ then follow:
\begin{eqnarray}
\frac{d}{dy}\left(X_{1}\frac{dR_{1}}{dy}\right)+\frac{X_{1}}{y}\frac{dR_{1}}{dy} -\frac{R_{1}}{X_{1}}\left(-y^{2}\Psi_{0}+\Psi_{1}-\frac{1}{y^{2}}\Psi_{2}\right)^{2} +\left(-b_{1}+\frac{1}{a_1^{2}a_2^{2}y^{2}}\Psi_{2}^{2}\right)R_{1}&=&0\;,\label{eqn:R1}\\
-\frac{d}{dr}\left(X_{2}\frac{dR_{2}}{dr}\right)-\frac{X_{2}}{r}\frac{dR_{2}}{dr}-\frac{R_{2}}{X_{2}}\left(r^{2}\Psi_{0}+\Psi_{1}+\frac{1}{r^{2}}\Psi_{2}\right)^{2}-\left(b_{1}+\frac{1}{a_1^{2}a_2^{2}y^{2}}\Psi_{2}^{2}\right)R_{2}&=&0\;,\label{eqn:R2}
\end{eqnarray}
where $b_{1}$ is the separation constant. 

\par By using the relationship between $t$, $\phi_{1}$, $\phi_{2}$ and $\psi_{0}$, $\psi_{1}$, $\psi_{2}$ in Eqs.~(\ref{eqn:20}-\ref{eqn:22}), we can obtain a relationship between the eigenvalues:
\begin{eqnarray}
\Psi_{0}\psi_{0}+\Psi_{1}\psi_{1}+\Psi_{2}\psi_{2} &=&-\omega t+m_1\phi_{1}+m_2\phi_{2}\nonumber\\
&=&[-\omega+g^{2}(m_1a_1+m_2a_2)]\psi_{0}+[-\omega(a_1^{2}+a_2^{2}) +m_1a_1(1+g^{2}a_2^{2})+m_2a_2(1+g^{2}a_1^{2})]\psi_{1}\nonumber\\ 
&&\ \ \ +[-\omega a_1^{2}a_2^{2}+m_1a_1a_2^{2}+m_2a_1^{2}a_2]\psi_{2}\;,
\end{eqnarray}
and
\begin{eqnarray}
-y^{2}\Psi_{0}+\Psi_{1}-\frac{1}{y^{2}}\Psi_{2} &=&\frac{1}{y^{2}}\left[\omega(y^{2}-a_1^{2})(y^{2}-a_2^{2})+ (1-g^{2}y^{2})\left(m_1a_1(y^{2}-a_2^{2})+m_2a_2(y^{2}-a_1^{2})\right)\right]\;.
\end{eqnarray}
We find then find:
\begin{eqnarray}
\label{angeq}
y\frac{d}{dy}\left[\left(\frac{1}{y}\right)(1-g^{2}y^{2})(a_{1}^{2}-y^{2}) (a_{2}^{2}-y^{2})\frac{dR_{y}}{dy}\right]+\left\{\left[a_{1}^{2}a_{2}^{2}-\frac{(a_{1}^{2}-y^{2})(a_{2}^{2}-y^{2})}{1-g^{2}y^{2}}\right]\omega^{2}
\right.&&\nonumber\\
+\left[\frac{a_{2}^{2}}{a_{1}^{2}}-\frac{(1-g^{2}y^{2})(a_{2}^{2}-y^{2})} {a_{1}^{2}-y^{2}}\right]m_{1}^{2}a_{1}^{2}+\left[\frac{a_{1}^{2}}{a_{2}^{2}}-\frac{(1-g^{2}y^{2})(a_{1}^{2}-y^{2})} {a_{2}^{2}-y^{2}}\right]m_{2}^{2}a_{2}^{2}\nonumber\\
+\left[ -b_1-2\omega(m_{1}a_{1}+m_{2}a_{2})+2g^{2}m_{1}a_{1}m_{2}a_{2}\right]y^{2}\bigg\}R_{y}&=&0\;,
\end{eqnarray}
with $0\leq\theta\leq\pi/2$ or $a_{2}\leq y\leq a_{1}$, where $b_1$ is a constant of separation. In the next section we shall effectively work with the variable $y$.

\par To be able to compare our spheroidal harmonic with other work in the literature, it is also interesting to write an expression for the angular wave equation in terms of the latitude variable $\theta$. From the angular equation in (\ref{eqn:R1}) above the derivative terms, with $y^{2}=a_1^{2}\cos^{2}\theta+a_2^{2}\sin^{2}\theta$, become
\begin{eqnarray}
\frac{d}{dy}\left(X_{1}\frac{dR_{1}}{dy}\right)+\frac{X_{1}}{y}\frac{dR_{1}}{dy} &=&\frac{1}{y}\frac{d}{dy}\left(yX_{1}\frac{dR_{1}}{dy}\right)\nonumber\\
&=&-\frac{1}{\sin\theta\cos\theta}\frac{d}{d\theta}\left[\left(1-g^{2}(a_1^{2}\cos^{2}\theta+a_2^{2}
\sin^{2}\theta)\right)\sin\theta\cos\theta\frac{dR_{1}}{d\theta}\right]\;.
\end{eqnarray}
The non-derivative term can be simplified to:
\begin{eqnarray}
&&-\frac{1}{X_{1}}\left(-y^{2}\Psi_{0}+\Psi_{1}-\frac{1}{y^{2}}\Psi_{2}\right)^{2} +\left(-b_{1}+\frac{1}{a_1^{2}a_2^{2}y^{2}}\Psi_{2}^{2}\right) \nonumber\\
&=&-b_{1}+\omega^{2}\left[\frac{1}{g^{2}}-\frac{(1-g^{2}a_1^{2})(1-g^{2}a_2^{2})} {g^{2}(1-g^{2}(a_1^{2}\cos^{2}\theta+a_2^{2}\sin^{2}\theta))}\right]+\omega\left[ -2(m_1a_1+m_2a_2)\right]\nonumber\\ 
&&\ \ \ +\left[g^{2}(m_1a_1+m_2a_2)^{2}+\frac{m_1^{2}}{\sin^{2}\theta}(1-g^{2}a_1^{2}) +\frac{m_2^{2}}{\cos^{2}\theta}(1-g^{2}a_2^{2})\right]\;.
\end{eqnarray}
Finally, the angular equation in terms of $\theta$ becomes
\begin{eqnarray}
&&\frac{d}{d\theta}\left[\left(1-g^{2}(a_1^{2}\cos^{2}\theta+a_2^{2} \sin^{2}\theta)\right)\sin\theta\cos\theta\frac{dR_{1}}{d\theta}\right]\nonumber\\
&&\ \ \ +\left[b_{1}-\frac{\omega^{2}}{g^{2}}+2\omega(m_1a_1+m_2a_2)-g^{2}(m_1a_1+m_2a_2)^{2}+\frac{\omega^{2}(1-g^{2}a_1^{2})(1-g^{2}a_2^{2})} {g^{2}(1-g^{2}(a_1^{2}\cos^{2}\theta+a_2^{2}\sin^{2}\theta))}\right. \nonumber\\
&&\ \ \ \ \ \ \left.-\frac{m_1^{2}}{\sin^{2}\theta}(1-g^{2}a_1^{2})-\frac{m_2^{2}}{\cos^{2}\theta}(1-g^{2}a_2^{2})\right]\sin\theta\cos\theta R_{1}=0\;.
\end{eqnarray}
This agrees with the angular equation in the general charged five-dimensional Kerr-(A)dS case with two rotations that was found in \cite{Aliev:2008yk}, also see reference \cite{Frolov:2002xf} for the flat $g=0$ case.  

\par For completeness, we also present the separation of the radial part, equation (\ref{eqn:R2}). We have
\begin{eqnarray}
-\frac{d}{dr}\left(X_{2}\frac{dR_{2}}{dr}\right)-\frac{X_{2}}{r}\frac{dR_{2}}{dr}
&=&-\frac{1}{r}\frac{d}{dr}\left[rX_{2}\frac{dR_{2}}{dr}\right]
=\frac{1}{r}\frac{d}{dr}\left(\frac{\Delta}{r}\frac{dR_{2}}{dr}\right)\;,
\end{eqnarray}
for the derivative terms, where we defined
\begin{equation}
\Delta=(1+g^{2}r^{2})(r^{2}+a_1^{2})(r^{2}+a_2^{2})-2Mr^{2}\;.
\end{equation}
In a similar way to the angular equation, we can simplify the non-derivative term in the radial equation as
\begin{eqnarray}
&&-\frac{1}{X_{2}}\left(r^{2}\Psi_{0}+\Psi_{1}+\frac{1}{r^{2}}\Psi_{2}\right)^{2} -\left(b_{1}+\frac{1}{a_1^{2}a_2^{2}r^{2}}\Psi_{2}^{2}\right) \nonumber\\
&=&-b_{1}+\omega^{2}\left[\frac{1}{g^{2}}-\frac{(1-g^{2}a_1^{2})(1-g^{2}a_2^{2})} {g^{2}(1+g^{2}r^{2})}\right]+\omega\left[ -2(m_1a_1+m_2a_2)\right]\nonumber\\ 
&&\ \ \ +\left[g^{2}(m_1a_1+m_2a_2)^{2}+\frac{m_1^{2}(a_1^{2}-a_2^{2})(1-g^{2}a_1^{2})}{r^{2}+a_1^{2}} +\frac{m_2^{2}(a_2^{2}-a_1^{2})(1-g^{2}a_2^{2})}{r^{2}+a_2^{2}}\right]\nonumber\\
&&\ \ \ +\frac{2M}{\Delta}(r^{2}+a_1^{2})(r^{2}+a_2^{2})(1+g^{2}r^{2}) \left[\frac{\omega}{1+g^{2}r^{2}}-\frac{m_1a_1}{r^{2}+a_1^{2}}-\frac{m_2a_2}{r^{2}+a_2^{2}} \right]^{2}\;.
\end{eqnarray}
The radial equation then becomes
\begin{eqnarray}
\frac{\Delta}{r}\frac{d}{dr}\left(\frac{\Delta}{r}\frac{dR_{2}}{dr}\right)+\left\{\Delta\left[ -b_{1}+\frac{\omega^{2}}{g^{2}}-2\omega(m_1a_1+m_2a_2)+g^{2}(m_1a_1+m_2a_2)^{2}\right.\right.&&\nonumber\\
\left.-\frac{\omega^{2}(1-g^{2}a_1^{2})(1-g^{2}a_2^{2})} {g^{2}(1+g^{2}r^{2})}+\frac{m_1^{2}(a_1^{2}-a_2^{2})(1-g^{2}a_1^{2})}{r^{2}+a_1^{2}} +\frac{m_2^{2}(a_2^{2}-a_1^{2})(1-g^{2}a_2^{2})}{r^{2}+a_2^{2}}\right]&&\nonumber\\
\left.+2M(r^{2}+a_1^{2})(r^{2}+a_2^{2})(1+g^{2}r^{2}) \left[\frac{\omega}{1+g^{2}r^{2}}-\frac{m_1a_1}{r^{2}+a_1^{2}}-\frac{m_2a_2}{r^{2}+a_2^{2}}\right]^{2}\right\}R_{2}&=&0\;.
\end{eqnarray}
This radial equation could be used to investigate QNMs, for example, on a five-dimensional Kerr-(A)dS background.


\section{Eigenvalue expansion for $D=5$}
\label{Ang5}

\par In Section \ref{sec:KG} we have given a general analysis of the massless scalar equation in a Kerr-(A)dS spacetime with two rotations and $D=5$ for general $g$. As we saw, the angular equation in this case has two rotation parameters, but only one angular equation, because there is only one latitude variable $\theta$ (cf. equation (\ref{lati})). This simplifies things a great deal and hence we shall now work out the perturbative expansion of the angular eigenvalue using techniques similar to the works of \cite{Fackerell:1977, Seidel:1988ue} (also see \cite{Mino:1997bw}). 

\par To develop a perturbative expansion for the eigenvalue $b_1$, it is convenient to make the change of variables
\begin{equation}
y^{2}=\frac{1}{2}\left(a_{1}^{2}+a_{2}^{2}\right)-\frac{1}{2}\left(a_{1}^{2}-a_{2}^{2}\right)x\;,
\end{equation}
with $-1\leq x\leq 1$. Then in terms of $x$, the angular equation becomes:
\begin{eqnarray}
&&-\left(1-x^2\right)\left[1-\frac{1}{2}g^{2}a_{1}^{2}(1-x)-\frac{1}{2}g^{2}a_{2}^{2}(1+x)\right]\frac{d^{2}R}{dx^{2}}\nonumber\\
&&\ \ +\left[2x-\frac{1}{2}g^{2}a_{1}^{2}(1-x)(1+3x)-\frac{1}{2}g^{2}a_{2}^{2}(1+x)(-1+3x)\right]
\frac{dR}{dx}\nonumber\\
&&\ \ \ +\frac{1}{4}\left[\frac{\omega^{2}}{g^{2}}-\frac{\omega^{2}(1-g^{2}a_{1}^{2})(1-g^{2}a_{2}^{2})}
{g^{2}\left(1-\frac{1}{2}g^{2}a_{1}^{2}(1-x)-\frac{1}{2}g^{2}a_{2}^{2}(1+x)\right)}
+\frac{2m_{1}^{2}(1-g^{2}a_{1}^{2})}{1+x}+\frac{2m_{2}^{2}(1-g^{2}a_{2}^{2})}{1-x}\right]R=\frac{B}{4}R\;,
\end{eqnarray}
where the constant $B=b_1+2\omega(m_{1}a_{1}+m_{2}a_{2})-g^{2}(m_{1}a_{1}+m_{2}a_{2})^{2}$. Note that for simplicity we have dropped the subscript on $R$.  

Now comes the perturbative part of the method, which is essentially identical to time-independent perturbation theory in quantum mechanics \cite{Mino:1997bw}. Expanding in powers of $a_{1}$ and $a_{2}$, which we assume to be small, we can schematically write
\begin{eqnarray}
&&\left({\cal O}_{0}+{\cal O}_{2}+{\cal O}_{4}+{\cal O}_{6}+\cdots\right)\left(R_{0}+R_{2}+R_{4}+R_{6}+\cdots\right)
\nonumber\\
&=&\frac{1}{4}\left(B_{0}+B_{2}+B_{4}+B_{6}+\cdots\right)\left(R_{0}+R_{2}+R_{4}+R_{6}+\cdots\right),\label{expansion}
\end{eqnarray}
where
\begin{eqnarray}
{\cal O}_{0}&=&-(1-x^2)\frac{d^{2}}{dx^{2}}+2x\frac{d}{dx}+\frac{1}{4}\left[\frac{2m_{1}^{2}}{1+x}+\frac{2m_{2}^{2}}{1-x}\right],\\
{\cal O}_{2}&=&\frac{1}{2}(1-x^2)\left[g^{2}a_{1}^{2}(1-x)+g^{2}a_{2}^{2}(1+x)\right]\frac{d^{2}}{dx^{2}} -\frac{1}{2}\left[g^{2}a_{1}^{2}(1-x)(1+3x)-g^{2}a_{2}^{2}(1+x)(1-3x)\right]\frac{d}{dx}\nonumber\\
&&\ \ +\frac{1}{4}\left[\frac{1}{2}\omega^{2}a_{1}^{2}(1+x)+\frac{1}{2}\omega^{2}a_{2}^{2}(1-x)
-\frac{2m_{1}^{2}g^{2}a_{1}^{2}}{1+x}-\frac{2m_{2}^{2}g^{2}a_{2}^{2}}{1-x}\right],\\
{\cal O}_{4}&=&\frac{1}{16}(a_{1}^{2}-a_{2}^{2})^{2}g^{2}\omega^{2}(1-x^{2}),\\
{\cal O}_{6}&=&\frac{1}{2}g^{2}\left[(a_{1}^{2}+a_{2}^{2})-(a_{1}^{2}-a_{2}^{2})x\right]{\cal O}_{4}\;.
\end{eqnarray}
The regularity of the form of the operators after ${\cal O}_{4}$ makes it possible to evaluate the eigenvalue iteratively as we shall do in the following.

\par We start with the zeroth order in equation (\ref{expansion}),
\begin{eqnarray}
{\cal O}_{0}R_{0}=\frac{1}{4}B_{0}R_{0},
\end{eqnarray}
which has regular solution
\begin{eqnarray}
\left(R_{0}\right)_{lm_{1}m_{2}}&=&c_{lm_{1}m_{2}}(1-x)^{|m_{2}|/2}(1+x)^{|m_{1}|/2}
P_{\frac{1}{2}\left(l-|m_{1}|-|m_{2}|\right)}^{(|m_{2}|,|m_{1}|)}(x)\\
\left(B_{0}\right)_{lm_{1}m_{2}}&=&l(l+2),
\end{eqnarray}
where $P_{n}^{(\alpha,\beta)}(x)$ is the Jacobi polynomial \cite{Gradshteyn:2000}. The normalization constant is
\begin{eqnarray}
c_{lm_{1}m_{2}}=\left\{\frac{(l+1)\Gamma\left[\frac{1}{2}(l-|m_{1}|-|m_{2}|)+1\right]
\Gamma\left[\frac{1}{2}(l+|m_{1}|+|m_{2}|)+1\right]}{2^{|m_{1}|+|m_{2}|-1}\Gamma\left[\frac{1}{2}(l+|m_{1}|-|m_{2}|)+1\right]
\Gamma\left[\frac{1}{2}(l-|m_{1}|+|m_{2}|)+1\right]}\right\}^{1/2},
\end{eqnarray}
with normalization 
\beq
\frac{1}{4}\int_{-1}^{1}dx\left(R_{0}\right)^{2}=1~.
\eeq 
Before moving on we will also need to consider how the normalization condition above affects the higher order terms:
\begin{eqnarray}
&&\frac{1}{4}\int_{-1}^{1}dx\left(R_{0}+R_{2}+R_{4}+R_{6}+\cdots\right)\left(R_{0}+R_{2}+R_{4}+R_{6}+\cdots\right)=1\nonumber\\
&\Rightarrow&\int_{-1}^{1}dx\ R_{0}R_{2}=0\ \ \ \ \ ;\ \ \ \ \ \int_{-1}^{1}dx\ R_{0}R_{4}=-\frac{1}{2}\int_{-1}^{1}dx\ R_{2}R_{2}\ \ \ \ \ ;\ \ \ \ \ \int_{-1}^{1}dx\ R_{0}R_{6}=-\int_{-1}^{1}dx\ R_{2}R_{4}\;.\label{orthoexp}
\end{eqnarray}

\par Now we consider the next order in equation (\ref{expansion}),
\begin{eqnarray}
{\cal O}_{2}R_{0}+{\cal O}_{0}R_{2}=\frac{1}{4}\left(B_{0}R_{2}+B_{2}R_{0}\right).\label{secondorder}
\end{eqnarray}
Contracting with $(R_{0})_{lm_{1}m_{2}}$ and making use of the fact that the operators are self-adjoint, we have
\begin{eqnarray}
(B_{2})_{lm_{1}m_{2}}=\int_{-1}^{1}dx\ \!(R_{0})_{lm_{1}m_{2}}{\cal O}_{2}(R_{0})_{lm_{1}m_{2}}\;.\label{B2}
\end{eqnarray}
From the properties of the Jacobi polynomial \cite{Gradshteyn:2000}, we obtain
\begin{eqnarray}
{\cal O}_{2}(R_{0})_{lm_{1}m_{2}}=(O_{2})_{l,l+2}\frac{c_{lm_{1}m_{2}}}{c_{(l+2)m_{1}m_{2}}}(R_{0})_{(l+2)m_{1}m_{2}}+(O_{2})_{l,l}(R_{0})_{lm_{1}m_{2}}+
(O_{2})_{l,l-2}\frac{c_{lm_{1}m_{2}}}{c_{(l-2)m_{1}m_{2}}}(R_{0})_{(l-2)m_{1}m_{2}}\;,\nonumber\\
\end{eqnarray}
with
\begin{eqnarray}
(O_{2})_{l,l+2}&=&\frac{1}{16}(a_{1}^{2}-a_{2}^{2})\left[\omega^{2}+g^{2}l(l+4)\right]\frac{[(l+2)^2-(|m_{1}|+|m_{2}|)^2]}{(l+2)(l+1)}\;,\\
(O_{2})_{l,l}&=&\frac{1}{8}(a_{1}^{2}+a_{2}^{2})\omega^{2}+\frac{1}{8}(a_{1}^{2}-a_{2}^{2})\omega^{2}\left[\frac{m_{1}^{2}-m_{2}^{2}}{l(l+2)}\right]
-\frac{1}{8}g^{2}(a_{1}^{2}+a_{2}^{2})l(l+2)
\nonumber\\
&&\ \ \ \ -\frac{1}{8}g^{2}(a_{1}^{2}-a_{2}^{2})(m_{1}^{2}-m_{2}^{2})\left[\frac{l^2+2l+4}{l(l+2)}\right],\\
(O_{2})_{l,l-2}&=&\frac{1}{16}(a_{1}^{2}-a_{2}^{2})\left[\omega^{2}+g^{2}(l^{2}-4)\right]
\frac{[l^2-(|m_{1}|-|m_{2}|)^2]}{l(l+1)}\;.
\end{eqnarray}
With this result, equation (\ref{B2}) gives
\begin{eqnarray}
(B_{2})_{lm_{1}m_{2}}&=&4(O_{2})_{l,l}\nonumber\\
&=&\frac{a_{1}^{2}+a_{2}^{2}}{2}\left[\omega^{2}-g^{2}l(l+2)\right]
+\frac{(a_{1}^{2}-a_{2}^{2})(m_{1}^{2}-m_{2}^{2})}{2l(l+2)}\left[\omega^{2}-g^{2}(l^{2}+2l+4)\right].
\end{eqnarray}

\par To go on to the next order we now need to express $(R_{2})_{lm_{1}m_{2}}$ in terms of the $(R_{0})_{l'm_{1}m_{2}}$ with the coefficients $(d_{2})_{ll'}$. Contracting equation (\ref{secondorder}) for $(R_{0})_{l'm_{1}m_{2}}$ with $l'\neq l$, we can obtain the coefficients $(d_{2})_{l,l'}$ in the following series expansion:
\begin{eqnarray}
(R_{2})_{lm_{1}m_{2}}=(d_{2})_{l,l+2}\frac{c_{lm_{1}m_{2}}}{c_{(l+2)m_{1}m_{2}}}(R_{0})_{(l+2)m_{1}m_{2}}
+(d_{2})_{l,l-2}\frac{c_{lm_{1}m_{2}}}{c_{(l-2)m_{1}m_{2}}}(R_{0})_{(l-2)m_{1}m_{2}}\;,
\end{eqnarray}
with
\begin{eqnarray}
(d_{2})_{l,l+2}&=&\frac{4}{(B_{0})_{lm_{1}m_{2}}-(B_{0})_{(l+2)m_{1}m_{2}}}(O_{2})_{l,l+2}\;,\\
(d_{2})_{l,l-2}&=&\frac{4}{(B_{0})_{lm_{1}m_{2}}-(B_{0})_{(l-2)m_{1}m_{2}}}(O_{2})_{l,l-2}\;.
\end{eqnarray}
From equation (\ref{orthoexp}), one has $(d_{2})_{l,l}=0$.

\par The next order in equation (\ref{expansion}) is
\begin{eqnarray}
{\cal O}_{4}R_{0}+{\cal O}_{2}R_{2}+{\cal O}_{0}R_{4}=\frac{1}{4}\left(B_{4}R_{0}+B_{2}R_{2}+B_{0}R_{4}\right),
\end{eqnarray}
where contracting with $(R_{0})_{lm_{1}m_{2}}$, we have
\begin{eqnarray}
(B_{4})_{lm_{1}m_{2}}=\int_{-1}^{1}dx\ \!\left[(R_{0})_{lm_{1}m_{2}}{\cal O}_{4}(R_{0})_{lm_{1}m_{2}}+(R_{0})_{lm_{1}m_{2}}{\cal O}_{2}(R_{2})_{lm_{1}m_{2}}\right]\;.\label{B4}
\end{eqnarray}
Thus, we need to consider the term ${\cal O}_{4}(R_{0})_{lm_{1}m_{2}}$, and to do that we use the Jacobi functional relation \cite{Gradshteyn:2000}:
\begin{eqnarray}
x \,(R_{0})_{lm_{1}m_{2}}=X_{l,l+2}\frac{c_{lm_{1}m_{2}}}{c_{(l+2)m_{1}m_{2}}}(R_{0})_{(l+2)m_{1}m_{2}}+X_{l,l}(R_{0})_{lm_{1}m_{2}}
+X_{l,l-2}\frac{c_{lm_{1}m_{2}}}{c_{(l-2)m_{1}m_{2}}}(R_{0})_{(l-2)m_{1}m_{2}}\;.
\end{eqnarray}
From the recurrence relation of the Jacobi polynomials, we have
\begin{eqnarray}
X_{l,l+2}&=&\frac{[(l+2)^{2}-(|m_{1}|+|m_{2}|)^{2}]}{2(l+1)(l+2)}\;,\nonumber\\
X_{l,l}&=&\frac{(m_{1}^{2}-m_{2}^{2})}{l(l+2)}\;,\nonumber\\
X_{l,l-2}&=&\frac{[l^{2}-(|m_{1}|-|m_{2}|)^{2}]}{2l(l+1)}\;.\label{xrecursion}
\end{eqnarray}
Then writing
\begin{eqnarray}
{\cal O}_{4}(R_{0})_{lm_{1}m_{2}}&=&(O_{4})_{l,l+4}\frac{c_{lm_{1}m_{2}}}{c_{(l+4)m_{1}m_{2}}}(R_{0})_{(l+4)m_{1}m_{2}}
+(O_{4})_{l,l+2}\frac{c_{lm_{1}m_{2}}}{c_{(l+2)m_{1}m_{2}}}(R_{0})_{(l+2)m_{1}m_{2}}
+(O_{4})_{l,l}(R_{0})_{lm_{1}m_{2}}\nonumber\\
&&\ \ +(O_{4})_{l,l-2}\frac{c_{lm_{1}m_{2}}}{c_{(l-2)m_{1}m_{2}}}(R_{0})_{(l-2)m_{1}m_{2}}
+(O_{4})_{l,l-4}\frac{c_{lm_{1}m_{2}}}{c_{(l-4)m_{1}m_{2}}}(R_{0})_{(l-4)m_{1}m_{2}}\;,
\end{eqnarray}
and using the result in equation (\ref{xrecursion}), we have
\begin{eqnarray}
(O_{4})_{l,l+4}&=&-\frac{1}{16}(a_{1}^{2}-a_{2}^{2})^{2}g^{2}\omega^{2}X_{l,l+2}X_{l+2,l+4}\;,\\
(O_{4})_{l,l+2}&=&-\frac{1}{16}(a_{1}^{2}-a_{2}^{2})^{2}g^{2}\omega^{2}(X_{l,l+2}X_{l+2,l+2}+X_{l,l}X_{l,l+2})\;,\\
(O_{4})_{l,l}&=&\frac{1}{16}(a_{1}^{2}-a_{2}^{2})^{2}g^{2}\omega^{2}(1-X_{l,l+2}X_{l+2,l}-X_{l,l}X_{l,l}-X_{l,l-2}X_{l-2,l})\;,\\
(O_{4})_{l,l-2}&=&-\frac{1}{16}(a_{1}^{2}-a_{2}^{2})^{2}g^{2}\omega^{2}(X_{l,l-2}X_{l-2,l-2}+X_{l,l}X_{l,l-2})\;,\\
(O_{4})_{l,l-4}&=&-\frac{1}{16}(a_{1}^{2}-a_{2}^{2})^{2}g^{2}\omega^{2}X_{l,l-2}X_{l-2,l-4}\;.
\end{eqnarray}
Putting these into equation (\ref{B4}), we have
\begin{eqnarray}
(B_{4})_{lm_{1}m_{2}}
&=&4\left[(O_{4})_{l,l}+(d_{2})_{l,l+2}(O_{2})_{l+2,l}+(d_{2})_{l,l-2}(O_{2})_{l-2,l}\right]\nonumber\\
&=&\frac{1}{64}(a_{1}^{2}-a_{2}^{2})^{2}\left\{\frac{8g^{2}\omega^{2}
[l^{4}+4l^{3}+2l^{2}(m_{1}^{2}+m_{2}^{2})+4l(m_{1}^{2}+m_{2}^{2}-2)-3(m_{1}^{2}-m_{2}^{2})^{2}]}{(l-1)l(l+2)(l+3)}\right.\nonumber\\
&&\ \ \ \ \ -\frac{[\omega^{2}+g^{2}l(l+4)]^{2}
[(l+2)^{2}-(|m_{1}|+|m_{2}|)^{2}][(l+2)^{2}-(|m_{1}|-|m_{2}|)^{2}]}{(l+1)(l+2)^{3}(l+3)}\nonumber\\
&&\ \ \ \ \ \left. +\frac{[\omega^{2}+g^{2}(l^{2}-4)]^{2}[l^{2}-(|m_{1}|+|m_{2}|)^{2}][l^{2}-(|m_{1}|-|m_{2}|)^{2}]}{l^{3}(l^{2}-1)}\right\}.
\end{eqnarray}

\par Using the same iterative procedure one can obtain the angular eigenvalue $B$ to higher orders in $a_{1}$ and $a_{2}$, where 
for completeness we present full expression up to 6th order in $B$:
\begin{eqnarray}
&&B=l(l+2)+\frac{a_{1}^{2}+a_{2}^{2}}{2}\left[\omega^{2}-g^{2}l(l+2)\right]
+\frac{(a_{1}^{2}-a_{2}^{2})(m_{1}^{2}-m_{2}^{2})}{2l(l+2)}\left[\omega^{2}-g^{2}(l^{2}+2l+4)\right]\nn
&+&\frac{1}{64}(a_{1}^{2}-a_{2}^{2})^{2}\left\{\frac{8g^{2}\omega^{2}
[l^{4}+4l^{3}+2l^{2}(m_{1}^{2}+m_{2}^{2})+4l(m_{1}^{2}+m_{2}^{2}-2)-3(m_{1}^{2}-m_{2}^{2})^{2}]}{(l-1)l(l+2)(l+3)}\right.\nn
&&\ \ \ \ \ -\frac{[\omega^{2}+g^{2}l(l+4)]^{2}
[(l+2)^{2}-(|m_{1}|+|m_{2}|)^{2}][(l+2)^{2}-(|m_{1}|-|m_{2}|)^{2}]}{(l+1)(l+2)^{3}(l+3)}\nn
&&\ \ \ \ \ \left. +\frac{[\omega^{2}+g^{2}(l^{2}-4)]^{2}[l^{2}-(|m_{1}|+|m_{2}|)^{2}][l^{2}-(|m_{1}|-|m_{2}|)^{2}]}{l^{3}(l^{2}-1)}\right\}\times\Big(1+ \frac{1}{2}g^{2}(a_{1}^{2}+a_{2}^{2}) \Big)\\
&+&\frac{1}{128}(a_{1}^{2}-a_{2}^{2})^{3}(m_{1}^{2}-m_{2}^{2})\times\nn
&&\ \ \left\{\frac{8g^{4}\omega^{2}
[l^{4}+4l^{3}-2l^{2}(3(m_{1}^{2}+m_{2}^{2})-4)-4l(3(m_{1}^{2}+m_{2}^{2})-2)+(5(m_{1}^{2}-m_{2}^{2})^{2}+8(m_{1}^{2}+m_{2}^{2})-16)]}
{(l^{2}-4)(l-1)l(l+3)(l+4)}\right.\nonumber\\
&&\ \ -\frac{[\omega^{4}-g^{2}\omega^{2}(3l^{2}+12l+20)-4g^{4}l(l+4)][\omega^{2}+g^{2}l(l+4)]
[(l+2)^{2}-(|m_{1}|+|m_{2}|)^{2}][(l+2)^{2}-(|m_{1}|-|m_{2}|)^{2}]}{l(l+1)(l+2)^{5}(l+3)(l+4)}\nn
&&\ \ \left. -\frac{[-\omega^{4}+g^{2}\omega^{2}(3l^{2}+8)+4g^{4}(l^{2}-4)][\omega^{2}+g^{2}(l^{2}-4)]
[l^{2}-(|m_{1}|+|m_{2}|)^{2}][l^{2}-(|m_{1}|-|m_{2}|)^{2}]}{l^{5}(l^{2}-1)(l^{2}-4)}\right\}\;.\nonumber
\label{mainB}
\end{eqnarray}
This is the main result of this paper,\footnote{A Mathematica notebook with results up to 8th higher order is available at \url{http://www-het.phys.sci.osaka-u.ac.jp/~naylor/AIM.html}.\label{weburl}} where some values are presented in Figure \ref{vary} andTable \ref{compar}. These are compared with an exact numerical procedure developed in the next section.

\par Note, this power series expansion of $B$ could be used to analyze the radial scalar perturbation equation for a general five-dimensional Kerr-(A)dS spacetime. As of yet only the case of $g=0$ has been considered for two rotations, e.g., see \cite{Frolov:2002xf,Nomura:2005mw} (however, see \cite{Aliev:2008yk}).

\section{The continued fraction method}
\label{res}

\par  Unlike the case for $D\geq 6$ \cite{Cho:2011yp}, the angular equation in five dimensions has four regular singular points. 
In this  situation, the equation can be transformed into Heun form and then the tried and tested method of continued fractions can be used \cite{Suzuki:1998vy}. Note, the asymptotic iteration method could also be applied \cite{Cho:2009wf,Cho:2011yp} to solve for the angular eigenvalue.

To get a continued fraction we first transform the angular equation (\ref{angeq}) according to 
\beq
R(\zeta)=\zeta^{\frac{|m_2|}{2}}(\zeta-1)^{\frac{|m_1|}{2}}(\zeta-z_0)^{\frac{\omega}{2g}}y(\zeta)
\eeq
where $\zeta=(\cos(2\theta)+1)/2$ and  
\bea
z_0&=&\frac{1-g^2a_2^2}{g^2(a_1^2-a_2^2)}.
\eea
Then the angular mode equation can be written in Heun form \cite{Suzuki:1998vy}:
\beq
\bigglb[\frac{d^2}{d\zeta^2}+\left(\frac{\gamma}{\zeta}+\frac{\delta}{\zeta-1}+\frac{\epsilon}{\zeta-z_0}\right)\frac{d}{d\zeta}+\frac{\alpha\beta \zeta-q}{\zeta(\zeta-1)(\zeta-z_0)}\biggrb]y(\zeta)=0 \,\,\, , 
\eeq
with the constraint 
\beq
\alpha+\beta+1=\gamma+\delta+\epsilon, \,\,\, 
\eeq
where
\bea
\alpha =\frac{1}{2} (|m_1| + |m_2| +\omega/g ), \quad\beta = \frac{1}{2} (|m_1| + |m_2|+ \omega/g)+2\quad\gamma=|m_2|+1, \quad\delta =|m_1|+1, \quad\epsilon = \omega/g+1,
\eea
and
\bea
q&=&\frac{1}{4g^4}\frac{\omega^2-Bg^2}{a_1^2-a_2^2}-\frac{m_1^2}{4}+\frac{1}{4}(m_2+\omega/g)(m_2+\omega/g+2)-\frac{1-g^2a_2^2}{4g^2(a_1^2-a_2^2)}(\omega^2/g^2-(m_1+m_2)(m_1+m_2+2)),\nn
\eea
\par A method identical to this was used by Kodama et al. \cite{Kodama:2009rq} for the case of a singly rotating black hole in AdS. A key point of the Heun form is that is satisfies a three term recurrence relation:
\bea
\alpha_0 c_{1}+\beta_0 c_0 &=& 0 \\
\alpha_p c_{p+1}+\beta_p c_p+\gamma_p c_{p-1}&=&0 \,,\qquad\qquad (p=1,2,\dots)\, ,
\eea
where
\bea
\alpha_p&=&-
\frac{(p + 1)(p + r - \alpha + 1)(p+ r -\beta + 1)(p + \delta)}
{(2p + r + 2)(2p + r + 1)}\,\,\,,\\
\beta_p&=&\frac{\epsilon p(p+r)(\gamma-\delta)+[p(p+r)+\alpha\beta][2p(p+r)+\gamma(r-1)]}{(2p + r + 1)(2p + r - 1)}-z_0p(p+r)-q\,\,\,,\\
\gamma_p&=&-
\frac{(p + \alpha - 1)(p + \beta - 1)(p + \gamma - 1)(p + r - 1)}
{(2p + r - 2)(2p + r - 1)}\,\,\,,
\eea
with
\beq
r  = |m_1|+|m_2|+1.
\eeq
The eigenvalue $B$ can then be found (for a given $\omega$) by solving a continued fraction of the form \cite{Leaver:1985ax, Berti:2005gp}:
\beq
\beta_0-\frac{\alpha_0\gamma_1}{\beta_1-}\frac{\alpha_1\gamma_2}{\beta_2-}\frac{\alpha_2\gamma_3}{\beta_3-}\ldots=0\,\,\,.
\label{CFM}
\eeq

\par We should finally mention that the limit $g= 0$ cannot be taken with the above method, because of the divergence in the exponent of equation (5.1). However, if the $g\to 0 $ limit is taken from the start it can be considered as a separate case, because it removes one of the regular singular points (from four to three) and reduces to a normal continued fraction solution. Incidentally, the asymptotic iteration method can be used even without the additional factor $(\zeta-z_0)^{\omega/2g}$ in the scaling and one can find the angular eigenvalues at $g= 0$ (see \cite{Cho:2009wf} for $a_2=0$).

\begin{figure}[h]
\centering
\scalebox{0.465}{\includegraphics{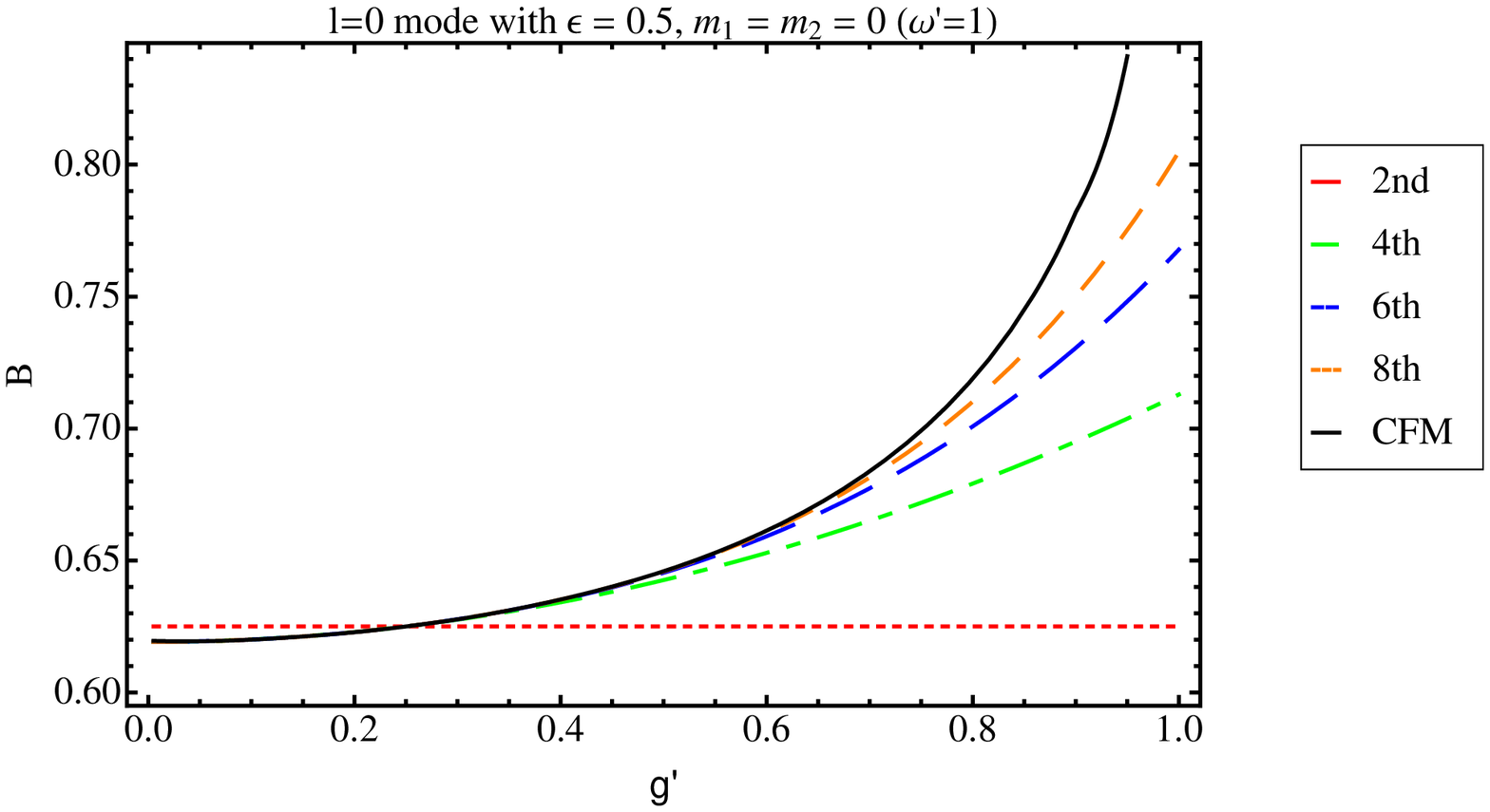}}
\hspace{0.0cm}
\scalebox{0.46}{\includegraphics{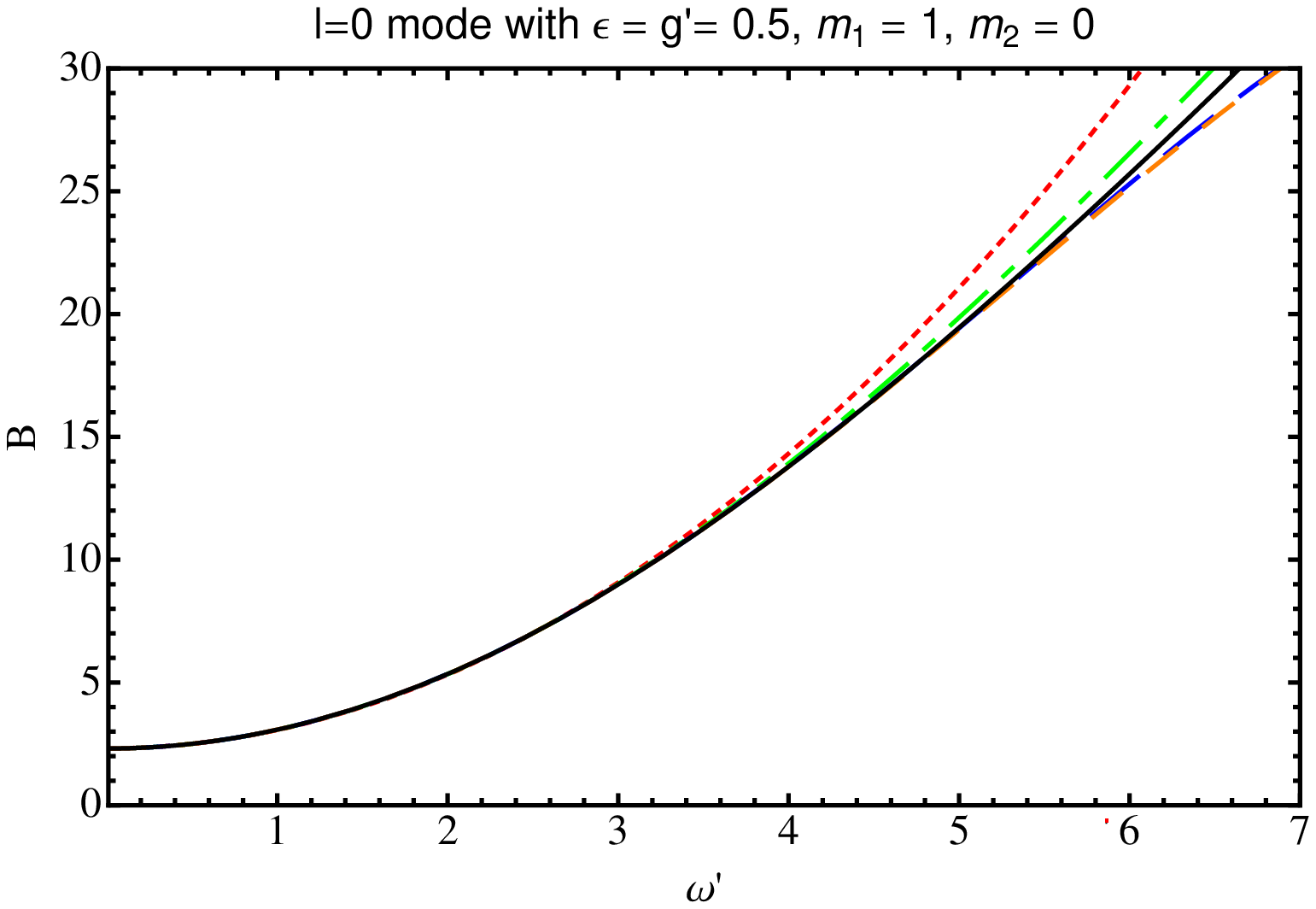}}
\caption{Comparison of the perturbative expansion up to 8th order with the continued fraction method (CFM) for $l=0$ modes. In the above plots $\epsilon = a_2/a_1$, $\omega^\prime = \omega a_1$ and $g^\prime=g a_1$  ($a_1=1.0$).}
\label{vary}
\end{figure}

\begin{table}[h]
\caption{\sl  Comparison between  the small $a_1,a_2$ expansion and the exact result for $l=0, m_1=1, m_2=0$ for given values of  $\epsilon=a_1/a_2$, $\omega^\prime=a_1 \omega = 1$ and $g^\prime=a_1 g= 0.01$. In the examples given below we found agreement between the numerical method and 6th \& 8th order accurate to $6 ~s.f.$}
\label{compar}
\begin{ruledtabular}
\begin{tabular}{ccccccc}
$(\epsilon,~ a_1)$ & (0.1, 0.5) &   (0.1, 1.0) & (0.5, 0.5)  &(0.5, 1.0) & (0.9, 0.5) &(0.9, 1.0)   \\
\hline
Numerical &   3.16715 & 3.66516 & 3.18727  &3.74711  & 3.23408 & 3.93621\\
\hline
2nd order  &  3.16743  &  3.66973 & 3.18743  & 3.74972  & 3.23409 & 3.93637  \\
\hline
4th order  & 3.16715   & 3.66521 &  3.18727 & 3.74711  & 3.23408 &  3.93621\\
\end{tabular}
\end{ruledtabular}
\end{table}

\section{Results \& Discussion}
\label{conc}

\par In summary we have separated the Klein-Gordon wave equation on a five-dimensional Kerr-(A)dS background with two rotation parameters and derived the radial and angular equations. The emphasis was on the scalar spheroidal harmonics, which are of relevance to charged and non-charged Kerr-(A)dS solutions \cite{Aliev:2008yk}. We developed perturbation theory and have presented the result for the angular eigenvalue, $B$, up to 6th order (see footnote \ref{weburl}). This now fills the gap, since a perturbative expansion for the angular eigenvalues in $D \geq 6$ has been studied in \cite{Cho:2011yp}.

\par In Figure \ref{vary} we show some plots for for varying $\omega^\prime, g^\prime$ for fixed values of $m_1$ and $m_2$. As expected we find improving agreement as the order of expansion increases. Note that in the left panel the 2nd order answer is only linear because for $l=m_1=m_2=0$ at 2nd order there is no $g$ dependence, cf. equation (\ref{mainB}). Table \ref{compar} also shows data for the $l=0, m_1=1, m_2=0$ mode, compared to the continued fraction method accurate to a precision of 6 s.f., which further confirms the improvement in increasing order.

\par It may be worth mentioning that we might also be able to form a perturbative expansion using invert continued fractions; this was successfully applied to simply rotating cases ($a_2=0$) in \cite{Berti:2005gp,Cho:2009wf}, where the natural choice appears to be in powers of $\omega^\prime=a_1 \omega$ and $\alpha_1 = a_1^2g^2$, cf. \cite{Cho:2009wf}. However, in the case of two rotation parameters there are clearly two choices: one is an expansion in powers of  $\omega^\prime=a_1 \omega$ and $\alpha_1 = a_1^2g^2, \alpha_2=a_2^2g^2$; the other (used in this paper)  is just an expansion in powers of $a_1$ and $a_2$. Preliminary work on an expansion of the inverted fraction for two rotations seems difficult to apply symbolically. However, it is possible to verify\footnote{H. T. Cho, unpublished notes.} that a perturbative expansion in terms of $\omega^\prime, \alpha_1,\alpha_2$ (albeit more cumbersome) does agree with the inverted fraction second order answer \cite{Berti:2005gp,Cho:2009wf} for a simply rotating black hole ($a_2=0$), when $a_2,\alpha_2\to 0$. We leave these interesting issues for future work.

\par In conclusion, given an analytic expression for the angular eigenvalue, this can now be used to find QNMs or absorption probabilities (greybody factors) for a massless scalar field separated on a Kerr-(A)dS background. As we mentioned in Section \ref{intro} this also applies to charged Kerr-(A)dS solutions (e.g., see \cite{Aliev:2008yk}) and although we have not shown it we expect that this is also true for the Kerr-(A)dS-NUT case. Interesting future work might be to look at QNMs for general five-dimensional solutions, particularly given their relevance to gauged supergravity models.



\bibliography{Kerr1stJune2011}{}

\end{document}